\documentclass[12pt]{article}
\usepackage{epsfig}
\usepackage{floatflt}
\usepackage{cite}

\def\Title#1{\begin{center} {\Large {\bf #1} } \end{center}}

\newcommand{\beq}[1]{$$ #1 $$}
\newcommand{\beqin}[1]{$ #1 $}

\newcommand{\bo}{\raise-1mm\hbox{\Large$\Box$}} % D'Alembertian

\newcommand{\bra}[1]{\langle #1|} \newcommand{\ket}[1]{|#1\rangle}

\newcommand{\eb}{\begin{equation}} \newcommand{\ee}{\end{equation}}

 \newcommand{\lt}{\left}
\newcommand{\rt}{\right}

\newcommand{\ds}{\displaystyle} \newcommand{\tev}{\,\mbox{TeV}}
\newcommand{\gev}{\,\mbox{GeV}} \newcommand{\mev}{\,\mbox{MeV}}

\newcommand{\Bbar}{\,\overline{\!B}}

\newcommand{\bbs}{$\mathrm{B_s}\!-\!\ov{\mathrm{B}}{}_\mathrm{s}\,$}
\newcommand{\bbms}{$\mathrm{B_s}\!-\!\ov{\mathrm{B}}{}_\mathrm{s}\,$\
mixing}
\newcommand{\bbmd}{$\mathrm{B_d}\!-\!\ov{\mathrm{B}}{}_\mathrm{d}\,$\
mixing}
\newcommand{\bbmq}{$\mathrm{B_q}\!-\!\ov{\mathrm{B}}{}_\mathrm{q}\,$\
mixing} 

\newcommand{\bbm}{$\mathrm{B}\!-\!\ov{\mathrm{B}}{}\,$\ mixing}
\newcommand{\kkm}{$\mathrm{K}\!-\!\ov{\mathrm{K}}{}\,$\ mixing}

\newcommand{\dm}{\ensuremath{\Delta m}}
\newcommand{\dg}{\ensuremath{\Delta \Gamma}}

\newcommand{\bea}{\begin{eqnarray}} \newcommand{\eea}{\end{eqnarray}}

\newcommand{\nn}{\nonumber \\} \newcommand{\no}{\nonumber}
\newcommand{\ov}{\overline} \newcommand{\epm}[2]{
\raisebox{-0.5ex}{\shortstack[l]{$\scriptstyle+#1$\\$\scriptstyle-#2$}}}
\newcommand{\fig}[1]{Fig.~\ref{#1}}

\newcommand{\eq}[1]{Eq.~(\ref{#1})}
\newcommand{\eqsand}[2]{Eqs.~(\ref{#1}) and (\ref{#2})}
\newcommand{\lqcd}{\Lambda_{\textit{\scriptsize{QCD}}}}

\def\journal#1#2#3#4{#1~{\bf #2}, #3 (#4)}

\def\PLB#1#2#3{\journal{Phys.\ Lett. B}{#1}{#2}{#3}}

\def\NPB#1#2#3{\journal{Nucl.\ Phys. B}{#1}{#2}{#3}}

\def\PRD#1#2#3{\journal{Phys.\ Rev. D}{#1}{#2}{#3}}

\newcommand{\arxiv}[1]{{arxiv:{#1}}}

\newcommand{\imag}{\mbox{Im}\,}

\newlength{\nseparation}
\begin{document}

\Title{B Mixing in the Standard Model and Beyond\footnote{Proceedings of
    CKM 2012, the 7th International Workshop on the CKM Unitarity
    Triangle, University of Cincinnati, USA, 28 September - 2 October
    2012 }}

\bigskip\bigskip

%+\addtocontents{toc}{{\it D. Reggiano}}
%+\label{ReggianoStart}

\begin{raggedright}  

{\it Ulrich Nierste\\
Institut f\"ur Theoretische Teilchenphysik\\ 
Karlsruhe Institute of Technology\\
Engesserstra\ss e 7\\
76131 Karlsruhe, Germany }
\bigskip\bigskip
\end{raggedright}

\begin{center}
\parbox[t]{0.8\textwidth}{
I present numerical updates of the Standard-Model predictions for the
mass and width differences and the CP asymmetries in flavor-specific
decays in \bbs\ and \bbmd.  Then I discuss the current status of new
physics in these mixing amplitudes.}
\end{center}

\boldmath
\section{\bbm : general formalism and $\dm$}
\unboldmath %%%%
\bbmq\ with \beqin{q=d} or \beqin{q=s} is governed by 
\beqin{M^q-i \Gamma^q/2} with the hermitian
\beqin{2\times 2} matrices \beqin{M^q} and \beqin{\Gamma^q}. The $(1,2)$
element of  \beqin{M^q-i \Gamma^q/2} induces \beqin{\Bbar_q \to B_q}
transitions. 

\begin{figure}[t]
\centering 
\epsfig{file=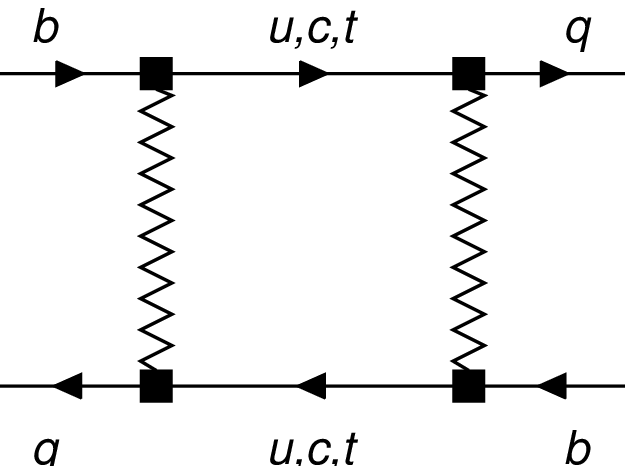,height=3.5cm} \hspace{3cm}
\epsfig{file=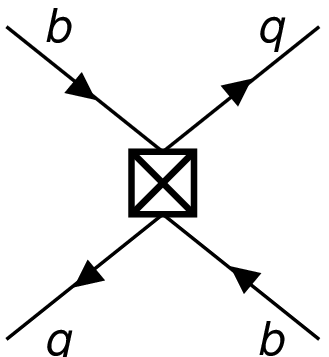,height=3.5cm}
\caption{Left: Box diagram. Right: Local operator $Q$.\label{fig:op}}
\end{figure}

The mass matrix element $M_{12}^q$ stems from the dispersive part of the
box diagram in \fig{fig:op}, which is obtained from the full diagram by
replacing the loop integral with its real part.  The decay matrix
element $\Gamma_{12}^q$ is calculated from the absorptive part of the box
diagram, which instead involves the imaginary part of the loop integral.
$M_{12}^q$ is dominated by the top contribution, while $\Gamma_{12}^q$
solely involves internal $c,u$ quarks and $|\Gamma_{12}^q|\ll
|M_{12}^q|$. \bbmq\ involves three
physical quantities:
\begin{eqnarray}
\lt| M_{12}^q \rt|,\quad  \lt| \Gamma_{12}^q \rt|,\quad
 \phi_q\equiv \arg \lt( - \frac{M_{12}^q}{\Gamma_{12}^q} \rt)
 \label{defphi}
\end{eqnarray}
The two eigenstates found by diagonalizing \beqin{M^q-i\,\Gamma^q/2}
differ in their masses and widths. A third observable is the CP
asymmetry in flavor-specific decays (usually called semileptonic CP
asymmetry), which quantifies CP violation in \bbmq.  The three
quantities in \eq{defphi} can be determined from the following 
observables:\\[2mm]
\centerline{
\begin{tabular}{l@{~~~~~~~~}l}
  \mbox{Mass difference:} &
  \beqin{\ds \dm_q  \simeq  2 |M_{12}^q|}, \\[1mm] 
  \mbox{Width difference:} &
  \beqin{\ds \dg_q \simeq  2 |\Gamma_{12}^q| \cos \phi_q}\\[1mm]
  \mbox{CP asymmetry in flavor-specific decays:} &
  \beqin{\ds a_{\rm fs}^q \simeq 
         \frac{|\Gamma_{12}^q|}{|M_{12}^{q}|} \sin \phi_q}
\end{tabular}}

$M_{12}^q$ is calculated with the help of an operator product expansion
(OPE), expressing this transition amplitude in terms of Wilson
coefficients and matrix elements of local four-quark operators.  The
Standard-Model prediction of $M_{12}^q$ only involves a single operator
$Q$:%
\bea%
M_{12} = (V_{tq}^* V_{tb})^2 \, C \, \bra{{\mathrm{B_q}}} Q
\ket{\ov{\mathrm{B}}{}_q} \label{m12}%
\eea%
Here \beqin{V_{tq}} and \beqin{V_{tb}} are the relevant elements of the
Cabibbo-Kobayashi-Maskawa (CKM) matrix.  The short-distance physics is
contained in $C=C(m_t,\alpha_s)$, which depends on the top quark mass
$m_t$ and the QCD coupling constant $\alpha_s$.  $C$ is known at the
level of next-to-leading-order corrections in QCD \cite{bjlw} and
suffers from very small theoretical uncertainties.
% \begin{nfloatingfigure}[r]{4cm}
% \centering 
% \epsfig{file=figs/db2-lo.eps,width=2.5cm}
% \caption{Operator.\label{fig:op}}
% \end{nfloatingfigure}
The four-quark operator $Q$ reads 
\begin{eqnarray}
 Q &=&  
    \ov{q}_L \gamma_{\nu} b_L \; \ov{q}_L \gamma^{\nu} b_L  \label{defq} 
\end{eqnarray}
and is depicted in \fig{fig:op}.

The theoretical uncertainty of \beqin{\dm_q } is dominated 
by the hadronic matrix element, which is parametrized as
\begin{eqnarray}
\bra{{\rm B_q}} Q \ket{\ov{\rm B}{}_q} & =& 
  \frac{2}{3} M_{B_q}^2 \, f_{B_q}^2 \, B_{B_q} . \label{me}
\end{eqnarray}
Here $ M_{B_q}$ and $f_{B_q}$ are mass and decay constant of $B_q$,
respectively, and $ B_{B_q} $ is called ``bag'' parameter. The matrix
element in \eq{me} is calculated with lattice QCD. The prediction of
$\dm_s$ involves $|V_{ts}|$, which is essentially equal to the
well-measured CKM element $|V_{cb}|$. With the result of
Ref.~\cite{bjlw} and present-day values of $V_{cb}$, $m_t$ and
$\alpha_s$ one finds %
\beq{\dm_s = \lt( \lt.\lt.\lt.  18.8 \pm
  0.6\rt._{V_{cb}} \pm 0.3\rt._{m_t} \pm 0.1\rt._{\alpha_s} \rt) \,
  \mbox{ps}^{-1}\, {\frac{f_{B_s}^2 \, B_{B_s} }{(220\mev)^2} }}%
with individual errors from the indicated sources.  Here the $\ov{\rm
  MS}$-NDR scheme for \beqin{B_{B_s}} is used and \beqin{B_{B_s}} is
evaluated at the scale \beqin{m_b}.  In the literature often the
scheme-invariant \beqin{\widehat{B}_{B_s}=1.51 B_{B_s}} is used instead.
%with an appropriately rescaled prefactor.

In phenomenological analyses usually also lattice results for the decay
constant $f_{B_s}$ are used. For instance, Ref.~\cite{Lenz:2012az} uses
the CKMfitter \cite{ckmfitter} averages of several lattice results,%
\bea%
f_{B_s}=(229\pm 2\pm 6) \mev, \qquad B_{B_s} = 0.85 \pm 0.02\pm 0.02. %
\label{lattnum}
\eea%
The quoted value for \beqin{B_{B_s}} is the average of the value in
Ref.~\cite{Aoki:2003xb} and of the value obtained from the ratio of
\beqin{f_{B_s}^2 \, B_{B_s}} and \beqin{f_{B_s}^2} calculated in
Ref.~\cite{Gamiz:2009ku}. 
With the numbers in \eq{lattnum} one finds 
\beqin{f_{B_s}^2 B_{B_s} = \lt[ (211 \pm 9) \mev \rt]^2}
and %
\bea%
\dm_s= (17.3 \pm 1.5 ) \, \mbox{ps}^{-1} % 
\label{dms}%
\eea%
complying excellently with the LHCb/CDF average \cite{hfag}%
\beq{\dm_s^{\rm exp}= (17.719 \pm 0.043 )\, \mbox{ps}^{-1}. }%
Bearing in mind that the prediction in \eq{dms} is essentially based on
calculations of $f_{B_s}$ rather than $f_{B_s}^2 B_{B_s}$ the quoted
error cannot be considered conservative. Using the preliminary lattice
result of the Fermilab/MILC collaboration \cite{Bouchard:2011xj},
\beqin{f_{B_s}^2 \, B_{B_s}=0.0559(68)\gev^2\simeq \lt[(237\pm 14)
  \mev\rt]^2}, one instead finds \beq{\dm_s= (21.7 \pm 2.6 ) \,
  \mbox{ps}^{-1} .} Clearly, new lattice results for \beqin{f_{B_s}^2
  B_{B_s}} are highly desirable to decrease the uncertainty in $\dm_s$ 
further. 

\begin{figure}[t]
\centering 
\epsfig{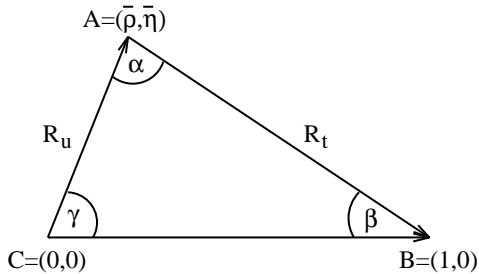} 
\caption{Unitarity triangle.\label{fig:ut}}
\end{figure}

Turning to \beqin{\dm_d}, I discuss the SM prediction for the ratio
\beqin{\dm_d/\dm_s }, from which \beqin{|V_{cb}|}, the short-distance
coefficient $C$ and some hadronic uncertainties drop out:
The hadronic quantity needed is 
\beq{\xi^2= \frac{f_{B_s}^2 B_{B_s}}{f_{B_d}^2 B_{B_d}}}
and the dependence on the CKM parameters reads:  %
 \bea%
\frac{\dm_d}{\dm_s} \propto \frac{|V_{td}|^2}{|V_{ts}|^2} 
     \propto R_t^2 %
\label{dmrt}%
\eea%
Here $R_t$ is one side of the unitarity triangle shown in \fig{fig:ut}.
The usual way to probe the Standard Model with \beqin{\dm_d} is to
perform a global fit to the unitarity triangle \cite{ckmfitter,utfit}. A
shortcut which reproduces the result of the global fit in an accurate
way exploits the calculation of $R_t$ from two angles of the UT
triangle: \beq{R_t = \frac{\sin \gamma}{\sin \alpha} = \frac{\sin
    (\alpha+\beta)}{\sin \alpha} } and the experimental data \cite{hfag}
\beq{\beta=21.4^\circ\pm 0.8^\circ,\qquad\quad \alpha=88.7^\circ
  \epm{4.6^\circ}{4.2^\circ}\;} give %
\bea%
R_t = 0.939 \pm 0.027. \label{rtang}%
\eea%
This number can be directly compared with the value of $R_t$ found from
$\dm_d/\dm_s$. Inverting the relation sketched in \eq{dmrt} yields
\cite[p.354]{run2}: %
\bea%
R_t &=& 0.880 \frac{\xi}{1.16} \sqrt{\frac{\dm_d}{0.49\,
    \mbox{ps}^{-1}}} \sqrt{\frac{17\, \mbox{ps}^{-1}}{\dm_s}}\,
\frac{0.22}{|V_{us}|} \,(1+ 0.050 \ov{\rho}) %
\eea%
With the Fermilab/MILC result \cite{Bazavov:2012zs} \beqin{\xi=1.268 \pm
  0.063} we find \beq{R_t=\lt.  0.942 \pm 0.047\rt._{\xi}\pm
  \lt. 0.006\rt._{\rm rest}} which agrees very well with
\eq{rtang}. Both numbers also agree with the CKMfitter global fit result
\beqin{R_t=0.926\epm{0.027}{0.028} } (using different lattice input)
obtained a few days before this conference.  Ignoring the small
deviation of $B_{B_s}/B_{B_d}$ from 1 the quantity $\xi$ equals the
ratio \beqin{f_{B_s}/f_{B_d}}, for which also a result obtained with QCD
sum rules is available, \beqin{\xi \simeq f_{B_s}/f_{B_d}=1.16\pm 0.04}
\cite{Jamin:2001fw}.  Data are now starting to challenge such low values
of $\xi$.

\boldmath
\section{Decay matrix \beqin{\Gamma_{12}}: 
           prediction of $\dg$ and $a_{\rm fs}$}
\begin{figure}[t]
\centering 
\epsfig{file=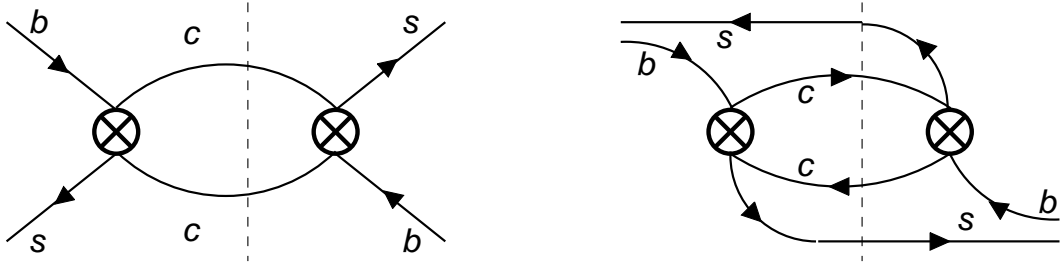,height=3.5cm} 
\caption{$\Gamma_{12}$ in the leading order of QCD.\label{fig:ga12}}
\end{figure}

\unboldmath%
\beqin{\Gamma_{12}^q}, \beqin{q=d,s}, is needed for the prediction of
the width difference \beqin{\ds \dg_q \simeq 2 |\Gamma_{12}^q| \cos
  \phi_q} and the semileptonic CP asymmetry \beqin{\ds a_{\rm fs}^q =
  \frac{|\Gamma_{12}^q|}{|M_{12}^{q}|} \sin \phi_q}.  For the
calculation of \beqin{\Gamma_{12}^q} another OPE, the so-called Heavy Quark
Expansion (HQE) is employed. The HQE expresses the $u,c$ contributions
to the \beqin{\Bbar_q \to B_q} transition amplitude in inverse powers of
$\lqcd/m_b$, the ratio of some hadronic scale $\lqcd\sim 500\,\mev$ and
the bottom quark mass.  The leading contribution to
\beqin{\Gamma_{12}^q} is depicted in \fig{fig:ga12}.
The HQE results in a simultaneous expansion of $\Gamma_{12}^q$ in
$\lqcd/m_b$ and $\alpha_s(m_b)$.  Corrections of order $\lqcd/m_b$ were
calculated in Ref.~\cite{bbd,Dighe:2001gc}, those of order $\alpha_s$
were obtained in Ref.~\cite{bbgln,bbln,rome}.  In Ref.~\cite{ln} these
NLO results have been expressed in terms of a new operator basis, which
leads to a better numerical stability by rendering an important
$\lqcd/m_b$ correction color-suppressed. Furthermore, in Ref.~\cite{ln}
the all-order re-summation of $\alpha_s^n z \ln^n z$ terms (with
$z=m_c^2/m_b^2$ and $n=1,2,\ldots$) (developed in Ref.~\cite{bbgln2}) has
been applied to $\Gamma^q_{12}$. To leading order in $\lqcd/m_b$ one
encounters two operators, $Q$ defined in \eq{defq} and 
\begin{eqnarray}
\widetilde{Q}_S &=& \ov{s}_L^\alpha b_R^\beta \, 
                    \ov{s}_L^\beta b_R^\alpha . \label{defqs}
\end{eqnarray}
with color indices $\alpha$ and $\beta$. The matrix element is
parametrized as 
\begin{eqnarray}
\bra{B_s} \widetilde{Q}_S 
\ket{\ov B_s} & = &  \frac{1}{12}
          M^2_{B_s}\, f^2_{B_s} \widetilde B_{S,B_s}^\prime.  \label{bst}
\end{eqnarray}
In the ratio \beqin{\dg_s/\dm_s} the dependence on \beqin{f_{B_s}^2
  B_{B_s}} drops out. We predict \beqin{\dg_s} by combining the theory
prediction of this quantity with \beqin{\dm_s^{\rm exp}= 17.719 \,
  \mbox{ps}^{-1}.}  For \beqin{m_t(m_t)=165.8\,\gev},
\beqin{m_b(m_b)=4.248\,\gev}, \beqin{m_c(m_c)=1.286\,\gev}, 
\beqin{m_s(m_b)=85\,\mev} and \beqin{\alpha_s(M_Z)=0.1184} 
(all in $\ov{\rm MS}$ scheme) one finds
\beq{\frac{\dg_s}{\dm_s} \dm_s^{\rm exp}= \lt[0.082\pm 0.007 + (0.019\pm
  0.001) \frac{\widetilde B_{S,B_s}^\prime }{B_{B_s}} - (0.027\pm 0.003)
  \frac{B_R}{B_{B_s}} \rt] \, \mbox{ps}^{-1}} Here \beqin{B_R} is a
generic bag parameter for the operators appearing at order $\lqcd/m_b$
and the quoted errors are the perturbative uncertainties estimated by
varying the renormalization scale.  All SM predictions presented in this
talk are an average of two renormalization schemes, using either the
$\ov{\rm MS}$ or pole definition for the overall factor of $m_b^2$
appearing in $\dg$ and $a_{\rm fs}$ (while re-summing $\alpha_s^n z \ln^n
z$ terms in both cases). Parametric errors (like those of the
quark masses) are of minor relevance.  

With the preliminary {Fermilab/MILC} result \cite{Bouchard:2011xj},
\beq{\frac{\widetilde B_{S,B_s}^\prime }{B_{B_s}}=1.50\pm 0.30,}
and the estimate $B_R=1\pm 0.5$ of the unknown higher-order bag
parameters one finds:
\begin{eqnarray} \frac{\dg_s}{\dm_s} \dm^{\rm exp} = \lt[0.078  \pm \lt. 
  0.016 \rt._{B_R/B} \pm \lt. 0.012
\rt._{\rm scale} \pm \lt. 0.008\rt._{\widetilde B/B} \rt] \,
\mbox{ps}^{-1}, \label{dgdmth}
\end{eqnarray}
which complies well with the LHCb measurement \cite{kyoto} 
\beq{\dg_s^{\rm LHCb} = \lt[0.116\pm 0.018{}_{\rm stat}\pm 0.006_{\rm syst}\rt]
  \mbox{ps}^{-1}}
and the LHCb/CDF/D\O\ average found by the HFAG \cite{hfag}:
\beq{\dg_s^{\rm exp} = \lt[ 0.089 \pm 0.012 \rt]  \mbox{ps}^{-1}.} 
For the same input as used in \eq{dgdmth} the CP asymmetry in
flavor-specific decays equals
\begin{eqnarray}  
  a_{\rm fs}^s &=& (1.8 \pm 0.3) \cdot 10^{-5}\label{afssth}
\end{eqnarray}
Here also the values of the CKM parameter are relevant, the quoted
number corresponds to \beqin{|V_{ub}|=3.49\cdot 10^{-3}}, 
\beqin{|V_{cb}|=40.89\cdot 10^{-3}}, and \beqin{\gamma=67.7^\circ} 
\cite{ckmfitter}.

In the \beqin{B_d} system the central value of \beqin{\dg/\dm} has
slightly shifted downwards from the 2006 value of \beqin{5.3\cdot
  10^{-3}} \cite{ln} to \beqin{\dg_d/\dm_d = 4.7\cdot 10^{-3}} with an
error of roughly 20\%. With \beqin{\dm_d^{\rm exp} =
  0.507\,\mbox{ps}^{-1}} this means \beqin{\dg_d=2.4\,\mbox{ns}^{-1}}
which is challenging to measure. The numerical prediction for the CP
asymmetry is
\begin{eqnarray}
  a_{\rm fs}^d &=& (-4.0 \pm 0.6)  \cdot  10^{-4}, \label{afsdth}
\end{eqnarray}
essentially unchanged from the update in \cite{ln11}. The quoted numbers 
use \beqin{\widetilde B_{S,B_d}^\prime/B_{B_d}=1.4\pm 0.4} inferred from 
\cite{Bouchard:2011xj}. Finally the CP phases in \eq{defphi} read 
\begin{eqnarray}
 \phi_s &=& 0.24^\circ \pm 0.06^\circ, \qquad\qquad   
 \phi_d \;=\; -4.9^\circ \pm 1.4^\circ. \label{phi} 
\end{eqnarray}

\section{New physics}
The D\O\ experiment has measured \cite{dimuon_evidence_d0,Abazov:2011yk}
\begin{eqnarray}
A_{\rm SL}^{\rm D0} &=& 
 (0.532 \pm 0.039) a_{\rm fs}^d + (0.468 \pm 0.039) a_{\rm fs}^s\nn
&=& (-7.87 \pm 1.72  \pm 0.93 ) \cdot 10^{-3} , \label{dzero}
\end{eqnarray}
which is $3.9\sigma$ off the SM prediction inferred from
\eqsand{afssth}{afsdth},
\begin{eqnarray}
  A_{\rm SL} &=& (-0.20 \pm 0.03) \cdot 10^{-3}. 
\end{eqnarray}
The prefactors of \beqin{ a_{\rm fs}^d} and \beqin{ a_{\rm fs}^s} in
\eq{dzero} are taken from Ref.~\cite{lnckmf12}, in which they were
calculated from the $B_{d,s}$ production fractions obtained by HFAG
\cite{hfag}. 

\begin{figure}[p,t]
\centering 
\epsfig{file=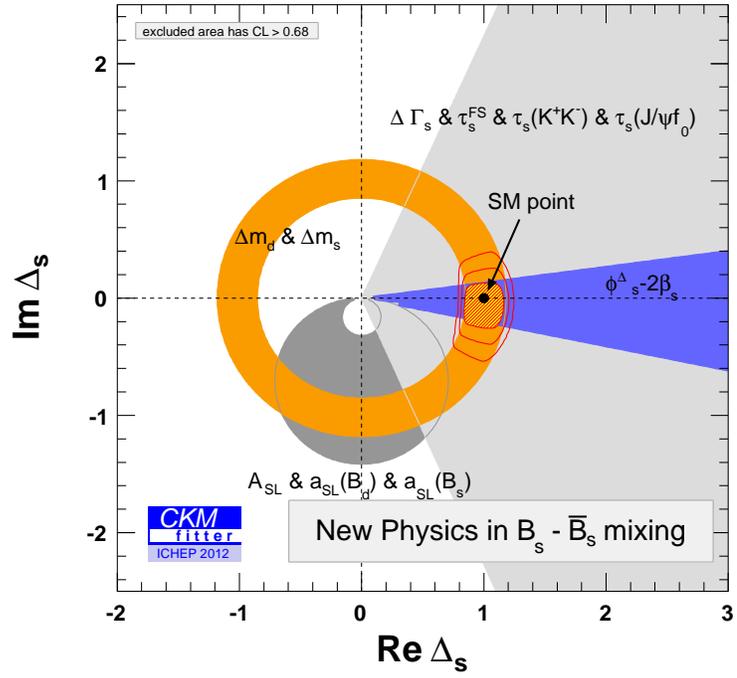,height=0.49\textheight} 
\epsfig{file=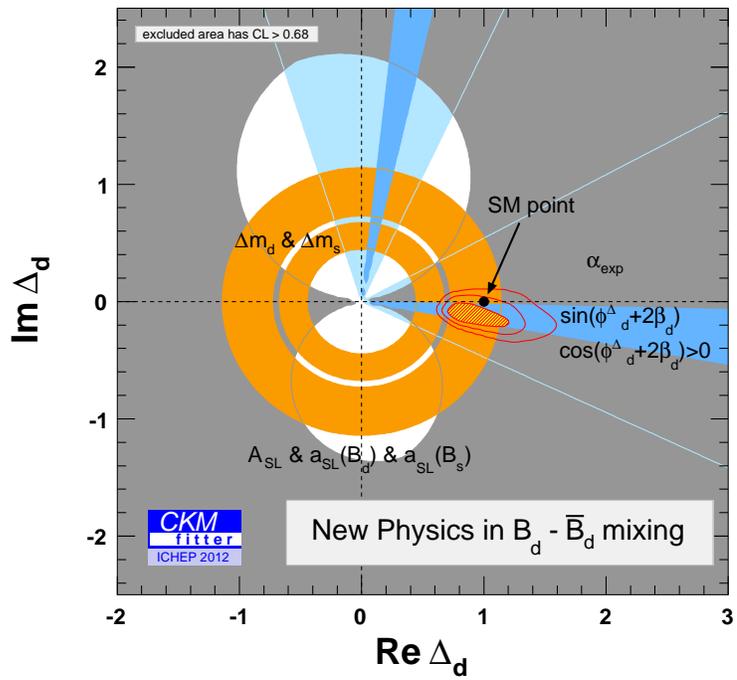,height=0.49\textheight} 
\caption{Allowed regions for the new-physics parameters \beqin{\Delta_s}
  and \beqin{\Delta_d} \cite{lnckmf12,ckmfitter}. \label{fig:s}}
\end{figure}

From a theoretical point of view, it is natural for new physics to
affect $M_{12}^s$ and $M_{12}^d$, which are sensitive to scales of 100$\,\tev$ and
  above. Parametrising the new-physics contribution as \cite{ln} 
\begin{eqnarray}
 M_{12}^q &  \equiv & M_{12}^{\rm SM,q} \cdot  \Delta_q \, ,
\qquad  \Delta_q \; \equiv \;  |\Delta_q| e^{i \phi^\Delta_q} ,
\qquad\quad q=d,s, \no
\end{eqnarray}
one can perform a global fit of \beqin{\Delta_{d,s}} and the CKM
elements to all relevant data.  This has been done in summer 2010
\cite{lnckmf} and spring 2012 \cite{lnckmf12}, before and after the
precise LHCb measurement of the CP violating phase in \beqin{B_s \to
  J/\psi \phi}, respectively. In 2010 the scenario with new physics in
\bbm\ and \kkm\ (and taking SM formulas for tree-dominated observables)
gave an excellent fit with a large, ${\cal O} (1)$ deviation of $\Delta
_s$ from its SM value $\Delta_s=1$ \cite{lnckmf}, with the SM point
\beqin{\Delta_d=\Delta_s=1} disfavored by 3.6$\sigma$. The new CP phase
\beqin{ \phi^\Delta_q} enters $a_{\rm fs}^q$ as \cite{ln}
\begin{eqnarray}
  a_{\rm fs}^s &=&  
   (4.4\pm 1.2 ) \cdot 10^{-3} 
\cdot \frac{\sin \left( \phi_s^{\rm SM} + \phi^\Delta_s
  \right)}{|\Delta_s|},\nn
  a_{\rm fs}^d & = &  
   (4.7\pm 1.4) \cdot 10^{-3} 
\cdot \frac{\sin \left( \phi_d^{\rm SM} + \phi^\Delta_d
  \right)}{|\Delta_d|} .  \label{afsnp}
\end{eqnarray}
The tagged analysis of \beqin{B_s\to J/\psi \phi} determines
\beqin{\phi_s^\Delta-2\beta_s} with $2\beta_s=2.1^\circ$. With the LHCb
data placing tight bounds on $|\phi_d^\Delta|$, the new-physics scenario 
described above cannot accommodate the D\O\ result anymore, the
SM point \beqin{\Delta_d=\Delta_s=1} is merely disfavored by 1$\sigma$.
The September-2012 update (see Ref.~\cite{ckmfitter}) of the plots 
presented in Ref.~\cite{lnckmf} are shown in Fig.~\ref{fig:s}. 
% and \ref{fig:d}. 
The pull value for \beqin{A_{\rm SL}} is found as $3.3\sigma$, showing
that the improvement compared to the SM is small. The fit prefers
$\phi_d^\Delta <0$ to loosen the tension with $A_{\rm SL}$ and to
accommodate the world average for $B(B\to \tau \nu)$
\cite{Lees:2012ju,Aubert:2009wt,Adachi:2012mm,Hara:2010dk}. This
branching ratio prefers a larger value of \beqin{|V_{ub}|} (despite of
the recent Belle result complying with the SM \cite{Adachi:2012mm}),
implying larger values of \beqin{R_u} (see \fig{fig:ut}) and
$\beta$. Since the CP asymmetry in \beqin{B_d\to J/\psi K_S} precisely
fixes \beqin{2\beta+\phi_d^\Delta=42.8^\circ \pm 1.6^\circ}, a larger
\beqin{|V_{ub}|} entails $\beta>21.4^\circ$ and therefore $\phi_d^\Delta
<0$ \cite{lnckmf,lnckmf12}. 

Note, however, that the 95\% CL region in the upper plot of \fig{fig:s}
is compatible with new physics in \beqin{M_{12}^s} of the order of
\beqin{30\%} of the SM amplitude. While the upcoming better LHCb data on
\beqin{B_s \to J/\psi \phi } will reduce this allowed region, these data
will only constrain \beqin{\phi_s^\Delta} and not \beqin{|\Delta_s|}. To
this end better lattice results for \beqin{f_{B_s}^2 B_{B_s} } are
urgently needed.

Several authors have considered the possibility of new physics in
$\Gamma_{12}^s$ from the yet unobserved decay mode \beqin{B_s \to
  \tau^+\tau^-} \cite{Dighe:2007gt,Bobeth:2011st,Dighe:2010nj}. The idea
of new physics in $B_s$ decays can reconcile the D\O\ result in
\eq{dzero} with the LHCb measurement of \beqin{\phi_s^\Delta-2\beta_s}
because \beqin{a_{\rm fs}^s = \imag (\Gamma_{12}^s/M_{12}^s)} grows with
new contributions to \beqin{\Gamma_{12}^s}. However, the LHCb experiment
has measured the average width \beqin{\Gamma_s} of the two $B_s$
eigenstates with high accuracy \cite{kyoto}. With the PDT value
\cite{pdt} for the $B_d$ width $\Gamma_d=1/\tau_{B_d}$ one finds
\beq{\frac{\Gamma_d}{\Gamma_s} = 0.997 \pm 0.013} in excellent agreement
with the SM prediction \beqin{\Gamma_d/\Gamma_s = 0.998 \pm 0.003}
\cite{bbd,kn,ln11}. This result precludes a sizable new $B_s$ decay 
rate into \beqin{\tau^+\tau^-} or other undetected final states
\cite{lnckmf,lnckmf12}.  However, phenomenologically,
sizable new physics in the doubly Cabibbo-suppressed quantity
\beqin{\Gamma_{12}^d} is still allowed \cite{lnckmf12}, but requires
somewhat contrived models of new physics.

\section{Conclusions} 
In this proceedings article I have updated several quantities related to
\bbs\ and \bbmd. It is stressed that the commonly used prediction of
$\dm_s$ relies on just two lattice calculations
\cite{Aoki:2003xb,Gamiz:2009ku}, which date back to 2003 and 2009. In
this article I have used the newer, but still preliminary results of the
Fermilab/MILC collaboration presented in
Refs.~\cite{Bazavov:2012zs,Bouchard:2011xj}. Significant numerical
differences with respect to the last update in Ref.~\cite{ln11} only
occur for \beqin{\dm_s}. 

The \bbms\ and \bbmd\ amplitudes are highly sensitive to new
physics. The LHCb measurements in \beqin{B_s \to J/\psi \phi} have
placed tight bounds on the CP phase in \beqin{M_{12}^s}, but
\beqin{{\cal O} (30\%)} new physics in \beqin{M_{12}^s} is still
possible. However, unlike in 2010 \cite{lnckmf} the hypothesis of new
physics only in \beqin{M_{12}^d} and \beqin{M_{12}^s} can no more
explain the D\O\ result for \beqin{A_{\rm SL}} \cite{lnckmf12}.

\bigskip
I am grateful to Alex Lenz for the opportunity to give this talk. The
presented work is supported by BMBF under grant no.~05H12VKF.

\end{document}